# Coupled Cluster Treatment of the Alternating Bond Diamond Chain

Jian-Jun Jiang [1], Yong-Jun Liu [2], Fei Tang [3], and Cui-Hong Yang[4]

[1]*Department of Physics, Sanjiang College, Nanjing 210012, China*

[2]*School of Physics Science and Technology, Yangzhou University, Yangzhou 225002, China*

[3]*Department of Electronic and Information Engineering, Yangzhou Polytechnic Institute, Yangzhou 225127, China*

[4]*Faculty of Mathematics and Physics, Nanjing University of Information Science and Technology, Nanjing 210044, China*

Abstract: By the analytical coupled cluster method (CCM), we study both the ground state and lowest-lying excited-state properties of the alternating bond diamond chain. The numerical exact diagonalization (ED) method is also applied to the chain to verify the accuracy of CCM results. The ED results show that the ground-state phase diagram contains two exact spin cluster solid ground states, namely, the tetramer-dimer (TD) state and dimer state, and the ferrimagnetic long-range-ordered state. We prove that the two exact spin cluster solid ground states can both be formed by CCM. Moreover, the exact spin gap in the TD state can be obtained by CCM. In the ferrimagnetic region, we find that the CCM results for some physical quantities, such as the ground-state energy, the sublattice magnetizations, and the antiferromagnetic gap, are comparable to the results obtained by numerical methods. The critical line dividing the TD state from the ferrimagnetic state is also given by CCM and is in perfect agreement with that determined by the ED method.

## 1. Introduction

Low-dimensional frustrated quantum spin systems have attracted continuous attention in recent years [1-10]. Owing to the interplay of strong quantum fluctuation in low dimensionality and frustration, low-dimensional frustrated quantum spin systems may possess not only the state with quasiclassical magnetic order, but also exotic quantum states with no classical counterpart, such as the spin-liquid state and valence-bond state [3, 8-10]. The impetus to the study of the properties of frustrated quantum spin systems was considerably enhanced after it was pointed out that there is a close connection between spin liquids and the high-Tc superconductivity of strongly correlated systems [11]. Although most low-dimensional frustrated quantum spin systems cannot exactly solved, there are still some exactly solvable ones, which are helpful for people to understand the nature of the exotic quantum phase [12-14]. The Majumdar-Ghosh model, which has been intensively studied, is a well-known example [12]. The properties of some quasi-one dimensional materials, such as $CuGeO_3$, can be described using that model [15]. The ground state (GS) of the Majumdar-Ghosh model is a spontaneous dimer state, which is the simplest type of spin cluster solid state that is a tensor product of the exact local eigenstates of cluster spins [16, 17]. Another prototypical spin chain that possesses the exact spin cluster solid state, which is different from that of the Majumdar-Ghosh model, is the pure spin-1/2 diamond chain [14]. The GS phase diagram of that chain is composed of the ferrimagnetic phase, dimer phase, and TD phase. The latter two phases both belong to the exact spin cluster solid state. Modifications of the diamond chain have attracted considerable attention in recent years, because of their rich spin cluster solid state [16-22]. In addition to the value of theoretical research, it is found that a variant of the diamond chain called the distorted diamond chain can be used to describe the magnetic lattice of the mineral azurite, $Cu_3(CO_3)_2(OH)_2$ [18].

Despite the fact that the diamond chain and its various modifications have the exact spin cluster solid GS in a certain parameter region, people often mainly used numerical methods, such as the ED method and density-matrix renormalization group (DMRG) method, to obtain the accurate GS phase diagram and lower-lying excited-state properties of those chains in previous research [14, 16-22]. Thus, it is an open question whether an analytical method can give the precise properties of the quantum diamond chain or its modifications. To answer the above question, we investigated in this work the properties of an alternating bond diamond chain (ABDC), a variant of the diamond chain, by an analytical method called CCM.

CCM is one of the most powerful methods of the quantum many-body theory [23]. Over the last twenty or so years, it has been applied with much success to various frustrated quantum spin systems in any dimension [8, 24-49]. Previous research shows that the results obtained by CCM are fully competitive with those obtained by other methods, such as series expansions and quantum Monte Carlo (QMC) [26, 29]. In this work, we aim to use CCM to investigate the properties of the ABDC. Our results indicate that CCM can yield accurate results for the GS and lower-lying excited state regardless of whether the GS of that chain is in the exotic quantum spin cluster state or ferrimagnetic state.

As shown in Fig. 1, the model Hamiltonian of the ABDC is

$$H = J_1\sum_{i=1}^{N/3} \vec{s}_{3i-2}\cdot(\vec{s}_{3i-1}+\vec{s}_{3i}) + J_2\sum_{i=1}^{N/3}\vec{s}_{3i-1}\cdot\vec{s}_{3i} + J_3\sum_{i=1}^{N/3}\vec{s}_{3i+1}\cdot(\vec{s}_{3i-1}+\vec{s}_{3i}),\qquad(1)$$

where $\vec{s}_{3i}$, $\vec{s}_{3i-1}$, and $\vec{s}_{3i-2}$ are spin-1/2 operators, and $J_1$, $J_2$, and $J_3$ are antiferromagnetic interactions. As the unit cell is made up of three sites, the total number of sites is $N$. For convenience, in what follows, we set $J_2 = \alpha J_1 = \alpha$ and $J_3 = \beta J_1 = \beta$. At $\beta = 0$, model (1) decouples into $N/3$ isolated triangles. The state of each spin triangle can be easily obtained. Thus, we will discuss the property of the Hamiltonian (1) at $\beta > 0$ in the following discussion.

The rest of the paper is arranged as follows. In Sect. 2, we determine the quantum phase diagram of the model by the ED method. In Sect. 3, CCM is used to discuss the property of the model. In Sect. 4, the CCM results are shown. A summary is given in Sect. 5.

**2. Quantum Phase Diagram of the Alternating Bond Diamond Chain**

To check the results of the CCM in the next section, we first give the quantum phase diagram of the ABDC by ED. Firstly, we consider some special cases of the Hamiltonian (1). The first one is the case of $\alpha = 0$. In this case, the Lieb-Mattis theorem implies that the magnitude of the total spin of the GS of the Hamiltonian (1), defined by $\vec{s}_{tol} = \sum_{l=1}^{N}\vec{s}_l$, is $N/6$ [50]. The ABDC possesses ferrimagnetic long-range order, and it is equivalent to an alternating bond antiferromagnetic mixed spin (1, 1/2) Heisenberg chain [51, 52]. It is reasonable to expect that the ferrimagnetic long-range order will extend to a small-$\alpha$-parameter region. In contrast, in the limiting case of $\alpha \to \infty$, the spins $\vec{s}_{3i-1}$ and $\vec{s}_{3i}$ form a singlet dimer and the GS is the disordered dimer state. As all spins $\vec{s}_{3i-2}$ are decoupled from each other in the dimer state, there is a $2^{N/3}$-fold degeneracy in that state. The GS energy per unit cell for the dimer state is given by

$$e_D = -0.75\alpha.\qquad(2)$$

The case of $\alpha = 1$ and $\beta = 1$ is the third one that we will discuss. The system reduces to a uniform diamond chain (UDC) in that case [14, 53]. The spin cluster state called the TD state is the exact GS of UDC when $\alpha = 1$ [14, 18]. As displayed in Fig. 2, the quadruplets $\vec{s}_{3i-2}$, $\vec{s}_{3i-1}$, $\vec{s}_{3i}$, and $\vec{s}_{3i+1}$ ($\vec{s}_{3i+1}$, $\vec{s}_{3i+2}$, $\vec{s}_{3i+3}$, and $\vec{s}_{3i+4}$) of spins form singlet tetramers, and the pairs $\vec{s}_{3i+2}$ and $\vec{s}_{3i+3}$ ($\vec{s}_{3i-1}$ and $\vec{s}_{3i}$) of spins construct singlet dimers in the TD state. Let us define the following composite spin operators: [16, 21, 53]

$$\vec{q}_i = \vec{s}_{3i-2} + \vec{s}_{3i-1} + \vec{s}_{3i} + \vec{s}_{3i+1} ,$$

$$\vec{t}_i = \vec{s}_{3i-1} + \vec{s}_{3i} . \tag{3}$$

Then, the TD state can be represented as

$$\left|\psi_{TD}^{\pm}\right\rangle = \frac{1}{\sqrt{2}}(\psi_{TD}^1 \pm \psi_{TD}^2) , \tag{4}$$

where $\psi_{TD}^1$ and $\psi_{TD}^2$ are degenerate and they have the forms

$$\left|\psi_{TD}^1\right\rangle = \prod_{i=1}^{N/6} \left|q_{2i-1} = 0, t_{2i-1} = 1, t_{2i} = 0\right\rangle$$

$$\left|\psi_{TD}^2\right\rangle = \prod_{i=1}^{N/6} \left|t_{2i-1} = 0, q_{2i} = 0, t_{2i} = 1\right\rangle \tag{5}$$

where

$$\left|q_{2i-1}=0,t_{2i-1}=1,t_{2i}=0\right\rangle = \frac{1}{\sqrt{12}}(-\left|\uparrow_{3i-2}\uparrow_{3i-1}\downarrow_{3i}\downarrow_{3i+1}\right\rangle - \left|\downarrow_{3i-2}\downarrow_{3i-1}\uparrow_{3i}\uparrow_{3i+1}\right\rangle - $$
$$\left|\uparrow_{3i-2}\downarrow_{3i-1}\uparrow_{3i}\downarrow_{3i+1}\right\rangle - \left|\downarrow_{3i-2}\uparrow_{3i-1}\downarrow_{3i}\uparrow_{3i+1}\right\rangle + 2\left|\downarrow_{3i-2}\uparrow_{3i-1}\uparrow_{3i}\downarrow_{3i+1}\right\rangle + 2\left|\uparrow_{3i-2}\downarrow_{3i-1}\downarrow_{3i}\uparrow_{3i+1}\right\rangle) \otimes$$
$$\frac{1}{\sqrt{2}}(\left|\uparrow_{3i+2}\downarrow_{3i+3}\right\rangle - \left|\downarrow_{3i+2}\uparrow_{3i+3}\right\rangle) ,$$

$$\left|t_{2i-1}=0,q_{2i}=0,t_{2i}=1\right\rangle = \frac{1}{\sqrt{2}}(\left|\uparrow_{3i-1}\downarrow_{3i}\right\rangle - \left|\downarrow_{3i-1}\uparrow_{3i}\right\rangle) \otimes \frac{1}{\sqrt{12}}(-\left|\uparrow_{3i+1}\uparrow_{3i+2}\downarrow_{3i+3}\downarrow_{3i+4}\right\rangle$$
$$-\left|\downarrow_{3i+1}\downarrow_{3i+2}\uparrow_{3i+3}\uparrow_{3i+4}\right\rangle - \left|\uparrow_{3i+1}\downarrow_{3i+2}\uparrow_{3i+3}\downarrow_{3i+4}\right\rangle - \left|\downarrow_{3i+1}\uparrow_{3i+2}\downarrow_{3i+3}\uparrow_{3i+4}\right\rangle$$
$$+2\left|\downarrow_{3i+1}\uparrow_{3i+2}\uparrow_{3i+3}\downarrow_{3i+4}\right\rangle + 2\left|\uparrow_{3i+1}\downarrow_{3i+2}\downarrow_{3i+3}\uparrow_{3i+4}\right\rangle) \tag{6}$$

where $\left|\uparrow\right\rangle$ and $\left|\downarrow\right\rangle$ are the $s^z$ eigenstates. If the TD state is the exact GS of the system, it is easy to obtain that the GS energy per unit cell is

$$e_{\text{TD}} = -0.5 - 0.25\alpha - 0.5\beta. \tag{7}$$

Compare Eq. (2) with Eq. (7), and you will find that the dimer state may be the GS of the Hamiltonian (1) only when $\alpha > 1 + \beta$. The case of $\beta \to \infty$ is the last special one that needs to be discussed. In that case, the three spins $\vec{s}_{3i-1}$, $\vec{s}_{3i}$, and $\vec{s}_{3i+1}$ form a trimer. The GS wave functions of the $i$th trimer are

$$\left|\phi_i^+\right\rangle = \frac{1}{\sqrt{6}}\left(\left|\uparrow_{3i-1}\downarrow_{3i}\uparrow_{3i+1}\right\rangle - 2\left|\uparrow_{3i-1}\uparrow_{3i}\downarrow_{3i+1}\right\rangle + \left|\downarrow_{3i-1}\uparrow_{3i}\uparrow_{3i+1}\right\rangle\right) \quad (s_{tol}^z = 1/2), \tag{8}$$

$$\left|\phi_i^-\right\rangle = \frac{1}{\sqrt{6}}\left(\left|\downarrow_{3i-1}\uparrow_{3i}\downarrow_{3i+1}\right\rangle - 2\left|\downarrow_{3i-1}\downarrow_{3i}\uparrow_{3i+1}\right\rangle + \left|\uparrow_{3i-1}\downarrow_{3i}\downarrow_{3i+1}\right\rangle\right) \quad (s_{tol}^z = -1/2), \tag{9}$$

where $s_{tol}^z$ is the $z$-component of the total spin of the $i$th trimer. Using the pseudo-operator $\vec{T}_i$ with the magnitude 1/2, one can express Eqs. (8) and (9) as [54, 55]

$$\begin{aligned}\left|\Uparrow_i\right\rangle &= \left|\phi_i^+\right\rangle \\ \left|\Downarrow_i\right\rangle &= \left|\phi_i^-\right\rangle\end{aligned}, \tag{10}$$

where $\left|\Uparrow_i\right\rangle$ and $\left|\Downarrow_i\right\rangle$ denote the eigenstate of $T_i^z$ with the eigenvalue 1/2 and the eigenstate of $T_i^z$ with the eigenvalue -1/2, respectively. In the fourth special case, the $J_1$ terms of the Hamiltonian (1) can be treated as perturbations. By using the first-order perturbation theory with respect to $J_1$, one can obtain the effective Hamiltonian

$$H_{\text{eff}} = -\frac{4}{9}J_1 \sum_{i=1}^{N/3} \vec{T}_i \cdot \vec{T}_{i+1}. \tag{11}$$

This result means that the GS of the Hamiltonian (1) is also in the ferrimagnetic state in the case of $\beta \to \infty$, just as in the first case discussed above.

Next, we determine the phase diagram by ED. As $\vec{t}_i^2$, defined by $\vec{t}_i^2 = t_i(t_i + 1)$, commutes with the Hamiltonian $H$, we have a sequence of good quantum numbers $\{t_1, t_2, \cdots t_i, \cdots t_{N/3}\}$. Thus, the GS of the ABDC belongs to one of the subspaces that are specified by $\{t_1, t_2, \cdots t_i, \cdots t_{N/3}\}$ [22]. As the magnitude of the composite spin $\vec{t}_i$ is 0 or 1, the correlation function between the spin pairs $\vec{s}_{3i-1}$ and $\vec{s}_{3i}$ takes a value of -0.75 or 0.25. One can then calculate the short-range correlation function $<\vec{s}_{3i-1} \cdot \vec{s}_{3i}>$ to determine the phase diagram of the ABDC. Our ED results show that the value of $<\vec{s}_{3i-1} \cdot \vec{s}_{3i}>$ is equal to 0.25, -0.25, or -0.75 in the entire parameter region. Thus, as shown in Fig. 3, the GS phase diagram of the ABDC is composed of the ferrimagnetic state, TD state, and dimer state. Finite-size effects on the position of the phase boundary are very minimal, as can be seen from the comparison of the results for $N$=12 and 30. At $\beta=1$, the ED results show that two critical points separate the TD state from the (i) ferrimagnetic state and (ii) dimer state. For a system with $N$=24, our results show that the two critical points are $\alpha=0.909$ and 2 respectively, which are consistent with those given in Ref. (14). Fig. 3 shows that, as expected above, the ABDC possesses the ferrimagnetic long-range order in the small-$\alpha$ or large-$\beta$-parameter region. Moreover, the ferrimagnetic state is always the GS of the chain if $\alpha$ is less than a certain critical value $\alpha_{TD}$. $\alpha_{TD}$ for a system with $N=30$ is shown in Fig. 3. When the parameter $\alpha$ exceeds that critical point, the TD phase appears in the phase diagram and it exists in a finite-parameter region. Besides $\alpha_{TD}$, the other critical value is $\alpha_D$ ($\alpha_D=1$), beyond which the dimer phase is also included in the phase diagram. The straight line $\beta=\alpha-1$ in Fig. 3 represents the exact boundary between the TD state and the dimer state.

## 3. Coupled Cluster Method Applied to the Alternating Bond Diamond Chain

In this section, we discuss the properties of the quantum TD state, dimer state, and ferrimagnetic state of the system determined by CCM. Since details of the CCM applied to quantum spin systems have been given elsewhere [23, 26, 27], we present only a brief description of the method that we used to treat the ABDC.

We first describe how we analyze the properties of the TD state by CCM. The starting and key point for a CCM calculation is to choose a suitable normalized reference state $|\phi\rangle$. In the past, people often chose the classical state or the quantum state (such as the dimer state) of the spin systems as the reference state of CCM to investigate the properties of the spin cluster state [8, 24]. Since the singlet tetramer and singlet dimer appear along the chain alternately in the TD state, we use the collinear state as shown in Fig. 1(a), but not the two types of state mentioned above, as the CCM reference state. As neighboring spins in the *A* and *B* sublattices are aligned parallel, whereas those in the *C* sublattice are aligned antiparallel, that reference state is also called the ferromagnetic-ferromagnetic-antiferromagnetic (FFA) state in the following discussion for convenience. After carrying out a mathematical rotation of the local axes of all the "up" spins: $s^x \to -s^x$, $s^y \to s^y$, and, $s^z \to -s^z$, all the spins in the reference state align along the negative *z*-axis. The reference state is then given by

$$|\phi\rangle = \cdots |\downarrow_{3i-2}\rangle \otimes |\downarrow_{3i-1}\rangle \otimes |\downarrow_{3i}\rangle \otimes |\downarrow_{3i+1}\rangle \otimes |\downarrow_{3i+2}\rangle \otimes |\downarrow_{3i+3}\rangle \otimes \cdots, \quad (12)$$

and the CCM parameterizations of the ket and bra GSs of model (1) are expressed as [26, 27]

$$|\psi\rangle = e^S |\phi\rangle, \quad S = \sum_{l=1}^{N} \sum_{i_1,i_2,\cdots i_l} S_{i_1,i_2,\cdots i_l} s^+_{i_1} s^+_{i_2} \cdots s^+_{i_l},$$

$$\langle \tilde{\psi}| = \langle \phi| \tilde{S} e^{-S}, \quad \tilde{S} = 1 + \sum_{l=1}^{N} \sum_{i_1,i_2,\cdots i_l} \tilde{S}_{i_1,i_2,\cdots i_l} s^-_{i_1} s^-_{i_2} \cdots s^-_{i_l}. \quad (13)$$

Because it is impossible to consider all the spin configurations in the $S$ and $\tilde{S}$ correlation operators in practice, we use the well-established LSUB*n* approximation scheme to truncate the expansions of $S$ and $\tilde{S}$ [26, 27]. Within the LSUB*n* approximation, only the configurations, including *n* or fewer correlated spins that span a range of no more than *n* contiguous lattice sites, are taken into account. In this paper, we assume that the two sites are contiguous if they are connected by $J_1$, $J_2$, or $J_3$ bonds. Although the number of fundamental configurations contained in the LSUB$n$ approximation grows rapidly with respect to the truncation index $n$, it can be reduced if we use the lattice symmetries and conservation laws that pertain to the Hamiltonian. Obviously, the LSUB$n$ approximation becomes exact in the limit $n \to \infty$.

Now, one can prove that the exact TD state of the Hamiltonian (1) can be produced by CCM with the FFA reference state. If all correlation coefficients contained in the ket-state correlation operator $S$ except those displayed in Fig. 4(a) are set equal to zero, $S$ is reduced to

$$S = S_4 \sum_{i=1}^{N/6} s^+_{3i-2} s^+_{3i-1} s^+_{3i} s^+_{3i+1} + S_2^a \sum_{i=1}^{N/6} (s^+_{3i-2} s^+_{3i-1} + s^+_{3i-2} s^+_{3i}) + S_2^b \sum_{i=1}^{N/6} (s^+_{3i-1} s^+_{3i+1} + s^+_{3i} s^+_{3i+1}) + S_2^c \sum_{i=1}^{N/6} s^+_{3i+2} s^+_{3i+3}. \quad (14)$$

The ket GS of model (1) is then given by

$$|\psi\rangle = e^S |\phi\rangle = \cdots \otimes [|\downarrow_{3i-2} \downarrow_{3i-1} \downarrow_{3i} \downarrow_{3i+1}\rangle + S_4 |\uparrow_{3i-2} \uparrow_{3i-1} \uparrow_{3i} \uparrow_{3i+1}\rangle + S_2^a |\uparrow_{3i-2} \uparrow_{3i-1} \downarrow_{3i} \downarrow_{3i+1}\rangle +$$
$$S_2^a |\uparrow_{3i-2} \downarrow_{3i-1} \uparrow_{3i} \downarrow_{3i+1}\rangle + S_2^b |\downarrow_{3i-2} \downarrow_{3i-1} \uparrow_{3i} \uparrow_{3i+1}\rangle + S_2^b |\downarrow_{3i-2} \uparrow_{3i-1} \downarrow_{3i} \uparrow_{3i+1}\rangle] \otimes \quad . \quad (15)$$
$$[|\downarrow_{3i+2} \downarrow_{3i+3}\rangle + S_2^c |\uparrow_{3i+2} \uparrow_{3i+3}\rangle] \otimes \cdots$$

By "re-rotating" the local axes of spins that point upward in the FFA reference state, one can obtain the following state: [8]

$$|\psi\rangle = \cdots \otimes [|\uparrow_{3i-2} \downarrow_{3i-1} \downarrow_{3i} \uparrow_{3i+1}\rangle + S_4 |\downarrow_{3i-2} \uparrow_{3i-1} \uparrow_{3i} \downarrow_{3i+1}\rangle - S_2^a |\downarrow_{3i-2} \uparrow_{3i-1} \downarrow_{3i} \uparrow_{3i+1}\rangle -$$
$$S_2^a |\downarrow_{3i-2} \downarrow_{3i-1} \uparrow_{3i} \uparrow_{3i+1}\rangle - S_2^b |\uparrow_{3i-2} \downarrow_{3i-1} \uparrow_{3i} \downarrow_{3i+1}\rangle - S_2^b |\uparrow_{3i-2} \uparrow_{3i-1} \downarrow_{3i} \downarrow_{3i+1}\rangle] \otimes \quad . \quad (16)$$
$$[|\downarrow_{3i+2} \uparrow_{3i+3}\rangle - S_2^c |\uparrow_{3i+2} \downarrow_{3i+3}\rangle] \otimes \cdots$$

It is obvious that an exact TD state is given by $S_4 = 1$, $S_2^a = S_2^b = 0.5$, and $S_2^c = 1$. Moreover, the GS energy per unit cell of the Hamiltonian (1) is written as

$$e_{TD} = \frac{3}{N} \langle \phi | e^{-S} H e^S | \phi \rangle = (-0.25 - 0.5 S_2^a) + \beta(-0.25 - 0.5 S_2^b) - 0.25 \alpha S_2^c = -0.5 - 0.25 \alpha - 0.5 \beta. \quad (17)$$

It is in agreement with Eq. (7).

Next, we prove that the exact dimer state can also be constructed within the CCM. To achieve this goal, the state shown in Fig. 1(b) is chosen to be the CCM reference state and we call it the ferromagnetic-ferromagnetic-ferromagnetic-I (FFFI) state. If only the ket-state correlation coefficient shown in Fig. 4(b) is not equal to zero, one can obtain the following ket-state within the CCM:

$$|\psi\rangle = \cdots \otimes |\downarrow_{3i-2}\rangle \otimes [|\downarrow_{3i-1}\uparrow_{3i}\rangle - S_2|\uparrow_{3i-1}\downarrow_{3i}\rangle] \otimes \cdots. \tag{18}$$

Apparently, the above state is only the dimer state if $S_2 = 1$. The GS energy per unit cell of the Hamiltonian (1) obtained by CCM based on the FFFI reference state is

$$e_D = \alpha(-0.25 - 0.5S_2) = -0.75\alpha. \tag{19}$$

It is obvious that Eq. (19) is the same as Eq. (2).

Although the above two short-range correlated spin cluster states are the exact GS of the Hamiltonian (1), the exact solution to the ferrimagnetic state cannot be obtained owing to quantum fluctuation. In the ferrimagnetic state, a pair of $\vec{s}_{3i-1}$ and $\vec{s}_{3i}$ forms a triplet dimer, and the magnitude of the total spin of that state is $N/6$ as mentioned above. Thus, we choose the state displayed in Fig. 1(c) as the CCM reference state to analyze the properties of the ferrimagnetic state. For convenience, the above reference state is also called the ferromagnetic-ferromagnetic-ferromagnetic-II (FFFII) state in the following discussion. We also calculate, aside from the GS energy, the typical physical quantity of the ferrimagnetic state, that is, the sublattice magnetizations $M_A$ and $M_{B+C}$ using

$$M_A = -\frac{1}{N/3}\sum_{i=1}^{N/3}\langle\tilde{\psi}|s^z_{3i-2}|\psi\rangle,$$

$$M_{B+C} = -\frac{1}{N/3}\sum_{i=1}^{N/3}\langle\tilde{\psi}|s^z_{3i-1}|\psi\rangle - \frac{1}{N/3}\sum_{i=1}^{N/3}\langle\tilde{\psi}|s^z_{3i}|\psi\rangle. \tag{20}$$

CCM can be well applied to investigating the properties of the lowest-lying excited state as well as the GS. The excited state wave function $|\psi_e\rangle$ is determined after applying an excitation operator $X^e$ linearly to the ket-state wave function. It is given by [26]

$$|\psi_e\rangle = X^e e^s|\phi\rangle, \quad X^e = \sum_{l=1}^{N}\sum_{i_1,i_2,\cdots i_l}\chi_{i_1,i_2,\cdots i_l}s^+_{i_1}s^+_{i_2}\cdots s^+_{i_l}. \tag{21}$$

Analogous to the GS, the LSUB$n$ approximation scheme is also used to truncate the expansion of the operator $X^e$. One can then use CCM to calculate the spin gap $\Delta$ of the spin systems. It is given by the lowest eigenvalue of the following LSUB$n$ eigenvalue equation: [26)]

$$\Delta \chi^e_{i_1,i_2,\cdots i_l} = \langle \phi | s^-_{i_1} s^-_{i_2} \cdots s^-_{i_l} e^{-S} [H, X^e] e^S | \phi \rangle \ . \tag{22}$$

We calculate two types of spin gap by CCM in the present paper. One is the single-triplet energy gap $\Delta_{ST}$, which is a representative physical quantity of the Hamiltonian (1) if its GS is the TD state. The spin gap $\Delta_{ST}$ is defined as

$$\Delta_{ST} = E_1(s_{tol} = 1) - E_g \ , \tag{23}$$

where $E_1$ and $E_g$ respectively denote the energy of the lowest-lying state with $s_{tol} = 1$ and the GS energy. In the ferrimagnetic phase, the ABDC possesses an antiferromagnetic character as well as a ferromagnetic character [51, 52)]. Thus, the other spin gap determined by CCM is the antiferromagnetic gap, which is given by

$$\Delta_{AF} = E_{N/6+1}(s_{tol} = N/6+1) - E_g, \tag{24}$$

where $E_{N/6+1}$ and $E_g$ are the energy of the lowest-lying state with $s_{tol} = N/6+1$ and the energy of the GS, respectively.

## 4. Results of the Coupled Cluster Method

Firstly, we present our CCM results for the GS. As TD and the dimer state are the exact GSs of the Hamiltonian (1), we focus on the properties of the ferrimagnetic state. When $\alpha = 0$ and $\beta = 1$, the property of the Hamiltonian (1) is exactly equivalent to that of the one-dimensional Heisenberg ferrimagnetic spin chain which has been investigated by various analytical and numerical methods [51, 56-58)]. Table I shows the results of CCM in that case. One can find that CCM results for the GS physical quantities, such as the GS energy per unit cell and the sublattice magnetization, converge very rapidly with an increase in the level of approximation. This phenomenon would be related to the short correlation length of the one-dimensional Heisenberg ferrimagnetic spin chain [58)]. As a result, the energy per unit cell $e$ and the sublattice magnetization $M_A$ given by CCM at the LSUB12 level of approximation are in agreement with four decimal places with the best results of the numerical method, namely, those of the DMRG method [56]. For clarity, we only show the results of the above physical quantities at the LSUB12 level of approximation in the following discussion [59]. To check the results of CCM, we also calculated those physical quantities by ED and found that the results obtained by ED also converge extremely fast. The physical quantities for a system with $N = 30$ are shown in Table I. It can be seen that they are very close to those of DMRG. Therefore, in the following part, the results of ED are also given for $N = 30$ sites for comparison with those of CCM.

Fig. 5 shows the GS energy $e$ as a function of $\beta$ given by CCM on the basis of the FFFII reference state for three distinct values of the parameter $\alpha$: $\alpha = 0$, $0.85$, and $1$. It can be found that the CCM results are in good agreement with those of ED in all cases. $e$ decreases monotonically with increasing in $\beta$ in the first case, whereas in the second (third) case, the energy determined by CCM on the basis of the FFFII reference state and that given by Eq. (7) intersect at two critical points (one point). This finding proves that the transition between the TD state and the ferrimagnetic state belongs to the first-order transition. By CCM, we have obtained the critical points at which the GS of the Hamiltonian (1) evolves from the ferrimagnetic state to the TD state for any other parameter $\alpha$ greater than $\alpha_{TD}$. They are presented in Fig. 6. One can see that the boundary line between the TD state and the ferrimagnetic state determined by CCM and that obtained by ED for a system with $N = 30$ almost overlap.

The results for the sublattice magnetizations $M_A$ and $M_{B+C}$ when $\alpha = 0$ are presented graphically in Fig. 7. As seen in that figure, the sublattice magnetization given by CCM coincides fairly well with that obtained by ED across the entire parameter range. $M_A$ and $M_{B+C}$ both experience growth with the increase in $\beta$ in the region $0 < \beta < 1$. At $\beta = 1$, they reach their maximum at the same time. Afterwards, they decrease with further increase in $\beta$. The reason for the evolution of sublattice magnetizations with the parameter $\beta$ is that the increase in $\beta - 1$ ($1 - \beta$) helps every three spins $\vec{s}_{3i-1}$, $\vec{s}_{3i}$, and $\vec{s}_{3i+1}$ ($\vec{s}_{3i-2}$, $\vec{s}_{3i-1}$, and $\vec{s}_{3i}$) form a trimer. As a result, the magnetic long-range order of the Hamiltonian (1) is strongest when $\beta = 1$.

Next, we present CCM results for the single-triplet energy gap $\Delta_{ST}$ and the antiferrimagnetic gap $\Delta_{AF}$, using $\alpha = 1$ as an example. In that case, whether the GS of the ABDC is in the TD state or ferrimagnetic state depends on whether the parameter $\beta$ is located in the region $0 < \beta < 1.18$ or $\beta > 1.18$. To check the results of CCM, $\Delta_{ST}$ and $\Delta_{AF}$ were also obtained by ED.

In Fig. 8, the ED results for $\Delta_{ST}$ are displayed when $0 < \beta < 1.18$. One can find that the single-triplet gap of a finite system with $N \geq 12$ reaches its value in the thermodynamic limit in the entire parameter region. Fig. 9 shows the single-triplet gap $\Delta_{ST}$ given by CCM. Apparently, CCM LSUB$n$ results for $\Delta_{ST}$ converge rapidly with an increase in $n$, and the spin gap in the limit $n \to \infty$ is determined by CCM if $n \geq 10$. Results of the spin gap $\Delta_{ST}$ in some cases are shown in Table II. As seen in Fig. 9 and Table II, the spin gap $\Delta_{ST}$ given by CCM with $n \geq 10$ equals that obtained by ED in the entire parameter region. The results of CCM and ED both show that the single-triplet gap obviously appears when $\beta > 0$, and increases with $\beta$ in the region $0 < \beta < 1$. When $\beta = 1$, it reaches its maximum. Although it then decreases with an increase in $\beta$, it does not vanish when $\beta < 1.18$. Hence, our current findings as well as the results of previous research indicate that the TD state is gapful [6, 60].

Finally, we turn to our CCM results for the antiferromagnetic gap $\Delta_{AF}$. The $\Delta_{AF}$

values for $\alpha = 0$ and $\beta = 1$ obtained from CCM are listed in Table I. One can see that our CCM results for $\Delta_{AF}$ are highly converged. The antiferromagnetic gap $\Delta_{AF}$ given by CCM at the LSUB12 level of approximation is in agreement up to three decimal places with that obtained by QMC [51]. The antiferromagnetic gap is plotted as a function of $\beta$ in Fig. 10 when $\alpha = 1$ and $\beta > 1.18$. $\Delta_{AF}$ values obtained by ED for a system with $N = 30$ are also displayed in that figure for comparison with the corresponding CCM data. One can find that $\Delta_{AF}$ increases with increasing in $\beta$, although the rate of increase gradually decreases. The size of the antiferromagnetic gap obtained by CCM is in good agreement with that given by ED in the entire parameter region. Thus, CCM can also be used to accurately analyze the lowest-lying excited-state properties of the ABDC.

## 5. Conclusions

In this paper, the CCM method, a powerful analytical tool for treating the frustrated Heisenberg chain in any dimension, was applied to the ABDC. To verify the accuracy of CCM results, we have also investigated the properties of the ABDC by the ED method. The ED results show that the GS phase diagram is composed of the TD state, dimer state, and ferrimagnetic state. We have shown that the former two exact spin cluster solid GSs can both be formed by CCM. Some physical quantities of the ferrimagnetic state, such as the GS energy and sublattice magnetizations, have been determined by CCM up to high orders of approximation. The results of the above quantities obtained by CCM are compared with those given by numerical methods. The case of $\alpha = 0$ and $\beta = 1$ is a typical example, in which the results of CCM are sufficiently accurate to be comparable to those of the numerical Monte Carlo or DMRG method. For any other parameter, the CCM results are also in perfect agreement with the results of the numerical method, namely, those of ED. Thus, it is natural to observe that CCM as well as ED can be used to precisely determine the phase boundary between the TD state and the ferrimagnetic state.

We have also calculated, aside from the GS physical quantities, the single-triplet energy gap and antiferromagnetic gap of the ABDC by CCM and compared them with those given by ED. Our results show that the single-triplet energy gap in the thermodynamic limit can be obtained by CCM. It is also found that the antiferromagnetic gap obtained by CCM is comparable to that determined by ED.

Therefore, the properties of the ABDC can be precisely analyzed by analytical CCM. Moreover, our findings provide a typical example of a powerful CCM application to frustrated quantum spin systems, even though its GS is in the quantum state with no classical analogy.

**Acknowledgments**

We thank Dr. Damain Farnell for his help with the application of CCM to spin systems. This work was supported by the Natural Science Foundation of Jiangsu Province (No. BK20131428) and the Natural Science Foundation of the Jiangsu Higher Education Institutions (No.13KJD140003).

Table I. Results obtained for the ABDC using the CCM in the case of $\alpha = 0$ and $\beta = 1$. The GS energy per unit cell $e$, the sublattice magnetization $M_A$, and the antiferrimagnetic gap $\Delta_{AF}$ obtained by CCM are shown. These results are compared with those obtained by other methods.

|  | $e$ | $M_A$ | $\Delta_{AF}$ |
|---|---|---|---|
| LSUB8 | -1.454172 | 0.292129 | 1.760226 |
| LSUB10 | -1.454109 | 0.292403 | 1.759433 |
| LSUB12 | -1.454096 | 0.292472 | 1.759224 |
| ED ($N$=30) | -1.454095 | 0.292478 | 1.759174 |
| Linear spin wave theory (SWT) [56] | -1.436 | 0.195 | 1 |
| Second-order SWT [57] | -1.454322 | 0.293884 | - |
| QMC [51, 58] | $-1.455 \pm 0.001$ | 0.29 | 1.75914 |
| DMRG [56] | $-1.45408$ | 0.29248 | - |

Table II. Results of the single-triplet energy gap using CCM-LSUB$n$ approximation with $n=\{8, 10, 12\}$ when $\alpha = 1$. These results are compared with those obtained by ED for $N$=30 sites.

| $\beta$ | LSUB8 | LSUB10 | LSUB12 | ED |
|---|---|---|---|---|
| 0.10 | 0.090396 | 0.080051 | 0.080051 | 0.080051 |
| 0.20 | 0.166259 | 0.146264 | 0.146264 | 0.146264 |
| 0.30 | 0.228444 | 0.198044 | 0.198044 | 0.198044 |
| 0.40 | 0.276621 | 0.235127 | 0.235127 | 0.235127 |
| 0.50 | 0.310449 | 0.257644 | 0.257644 | 0.257644 |
| 0.60 | 0.329849 | 0.266132 | 0.266132 | 0.266132 |
| 0.70 | 0.335139 | 0.261484 | 0.261484 | 0.261484 |
| 0.80 | 0.327065 | 0.244852 | 0.244852 | 0.244852 |
| 0.90 | 0.306736 | 0.217528 | 0.217528 | 0.217528 |
| 1.00 | 0.275494 | 0.180828 | 0.180828 | 0.180828 |
| 1.10 | 0.234759 | 0.136015 | 0.136015 | 0.136015 |
| 1.15 | 0.211265 | 0.110931 | 0.110931 | 0.110931 |

# Figure captions

Fig. 1. Sketches of the FFA reference state (a), FFFI reference state (b), and FFFII reference state (c) of the ABDC.

Fig. 2. Schematic picture of the TD state. The rectangles and ellipses represent the tetramers and singlet dimers, respectively.

Fig. 3. GS phase diagram obtained by ED. The solid line $\beta = \alpha - 1$ represents the exact boundary between the dimer phase and the TD phase.

Fig. 4. Illustration of fundamental configurations retained in the ket-state correlation operator $S$ for CCM based on FFA reference state (a) or FFFI reference state (b). The centers of the shaded circles mark the flipped spins with respect to the reference state.

Fig. 5. GS energy per site $e$ versus $\beta$ using CCM based on FFFII reference state, ED, and Eq. (7) for different $\alpha$ values.

Fig. 6. Boundary line between the TD state and the ferrimagnetic state determined by CCM and ED.

Fig. 7. Sublattice magnetizations $M_A$ and $M_{B+C}$ versus $\beta$ using CCM based on FFFII reference state and ED when $\alpha = 0$.

Fig. 8. Spin gap $\Delta_{ST}$ of the ABDC versus $\beta$ using ED when $\alpha = 1$.

Fig. 9. Spin gap $\Delta_{ST}$ of the ABDC versus $\beta$ using CCM based on FFA reference state and ED when $\alpha = 1$.

Fig. 10. Spin gap $\Delta_{AF}$ versus $\beta$ using CCM based on FFFII reference state and ED when $\alpha = 1$.

**Figure 1**

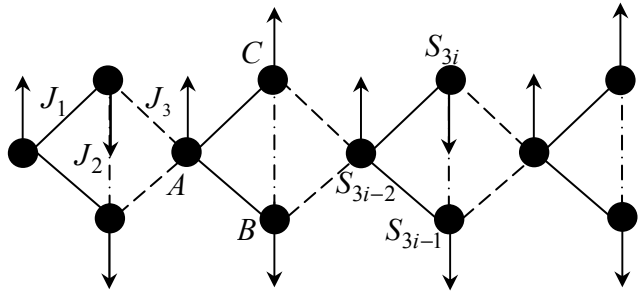

**(a)**

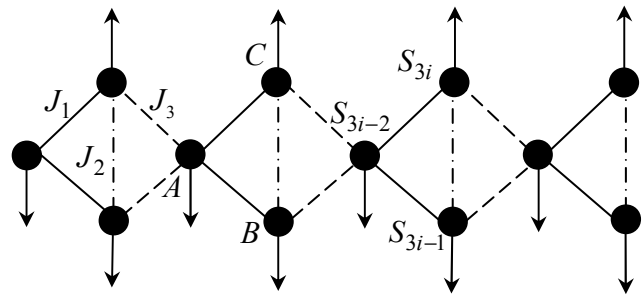

**(b)**

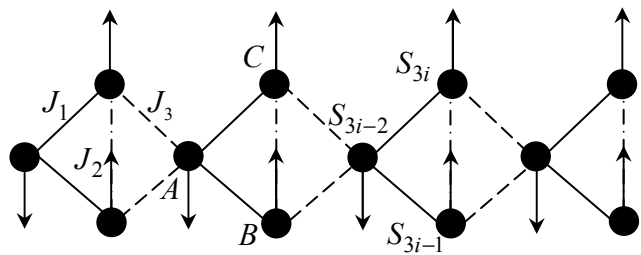

**(c)**

**Figure 2**

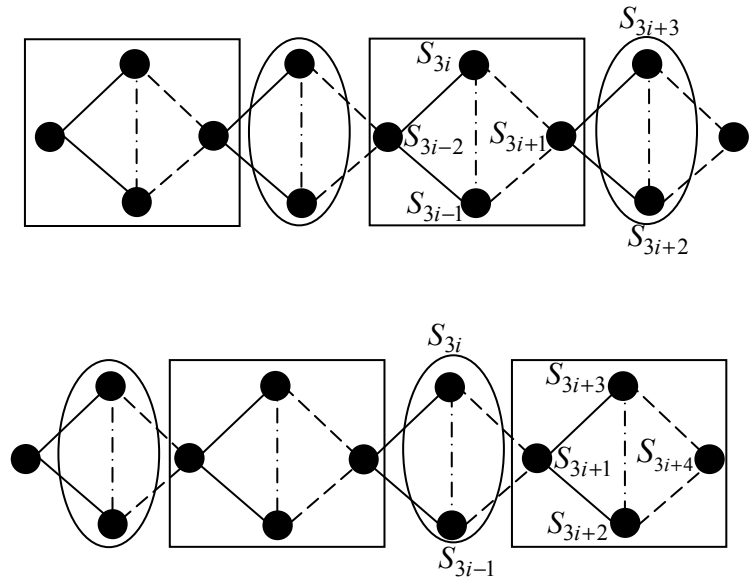

**Figure 3**

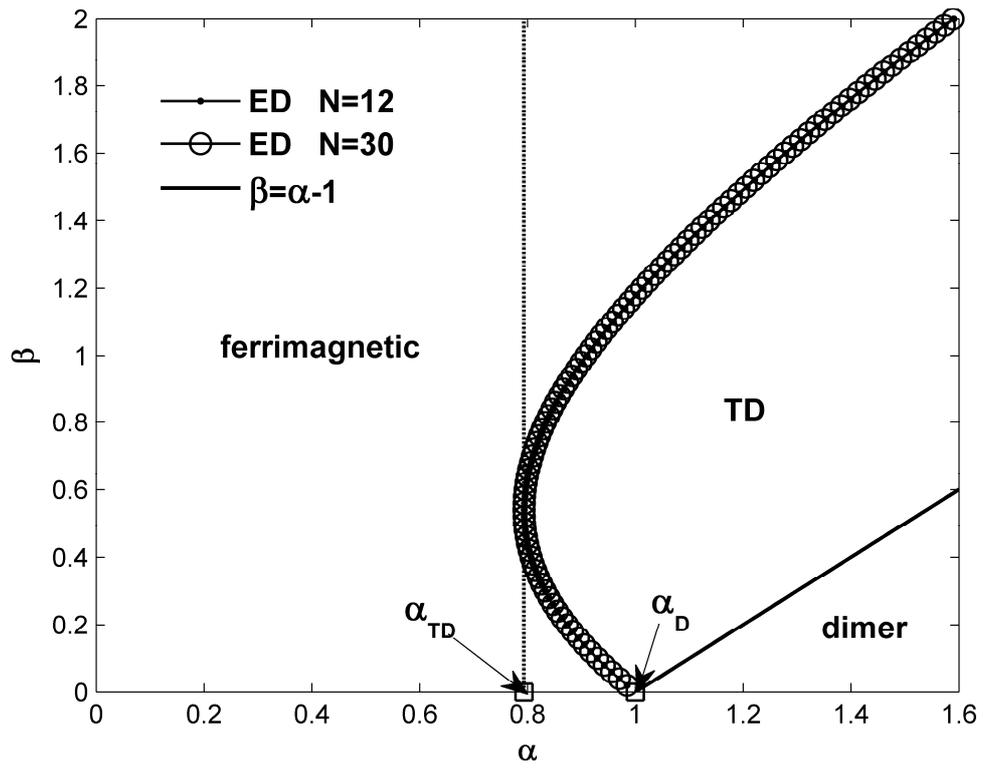

**Figure 4**

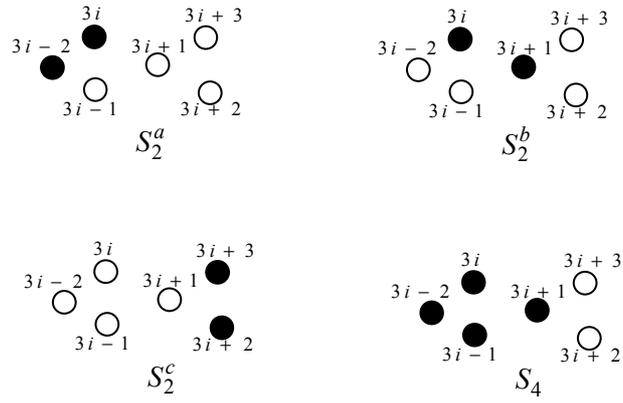

**(a)**

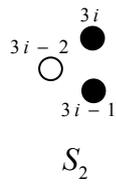

**(b)**

**Figure 5**

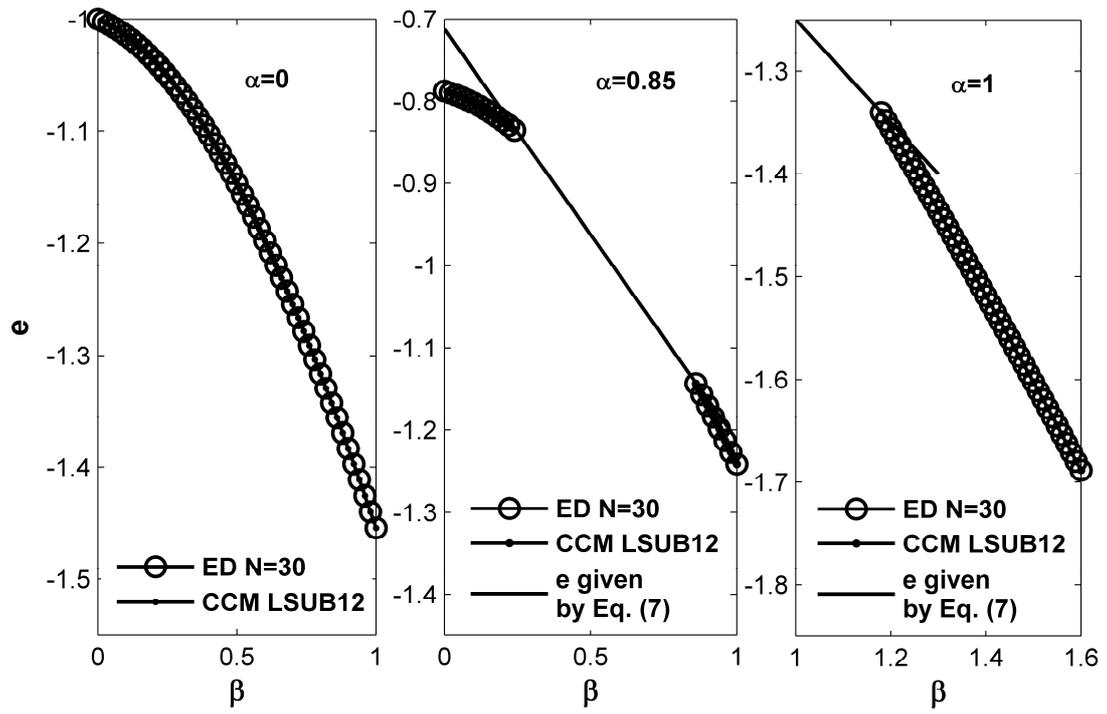

**Figure 6**

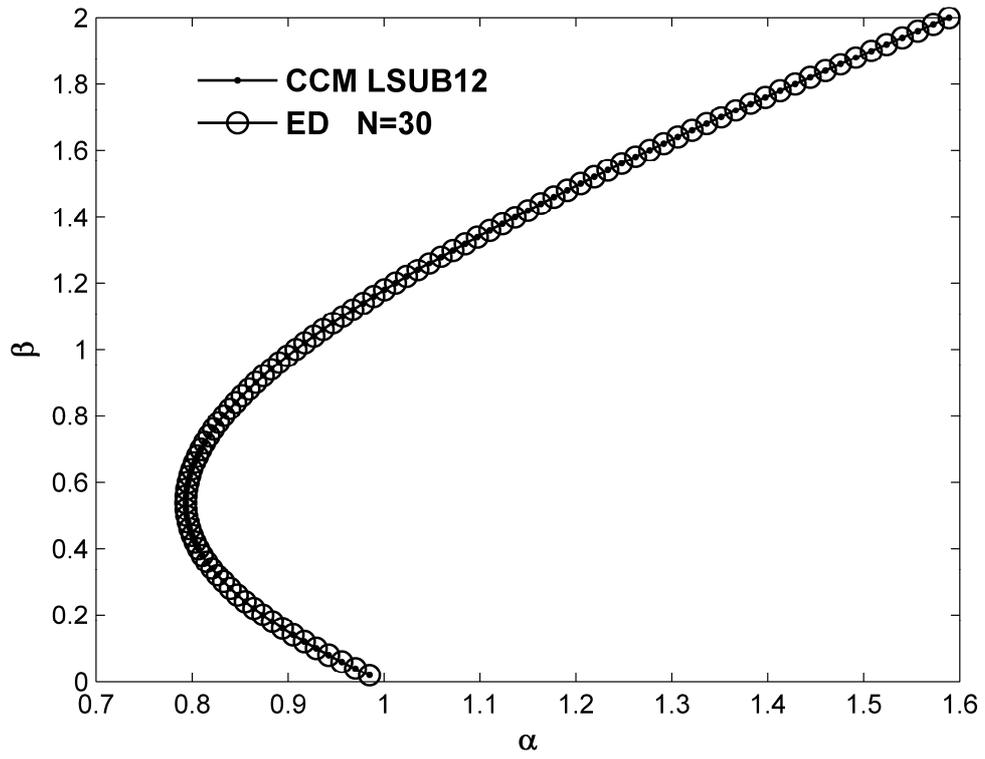

**Figure 7**

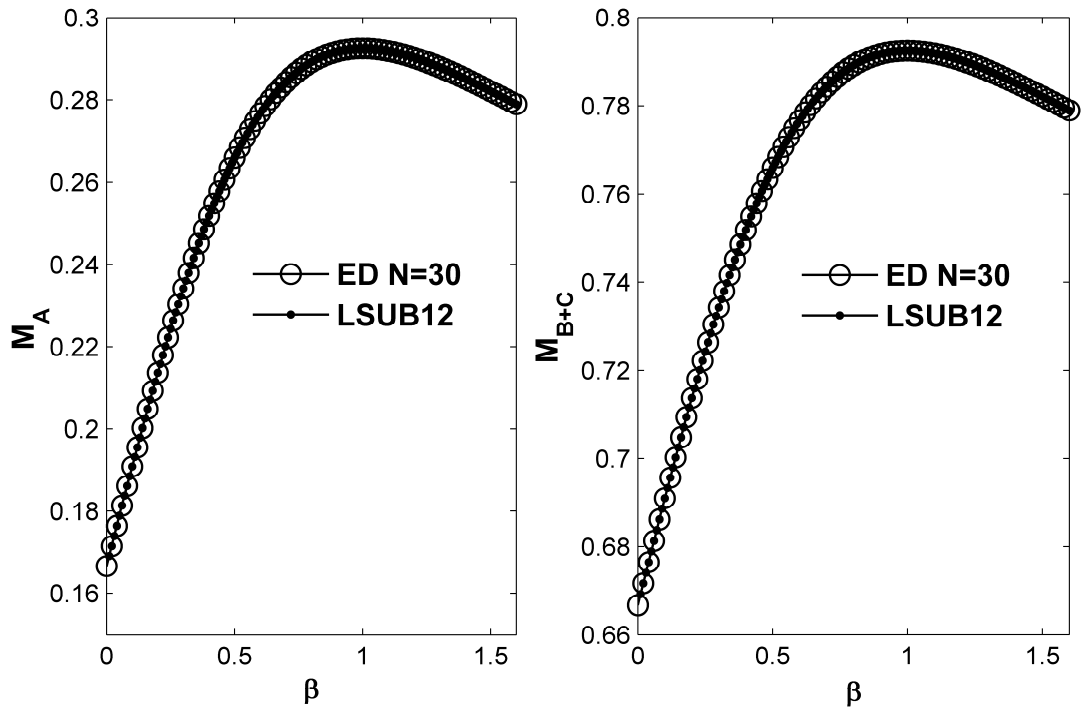

**Figure 8**

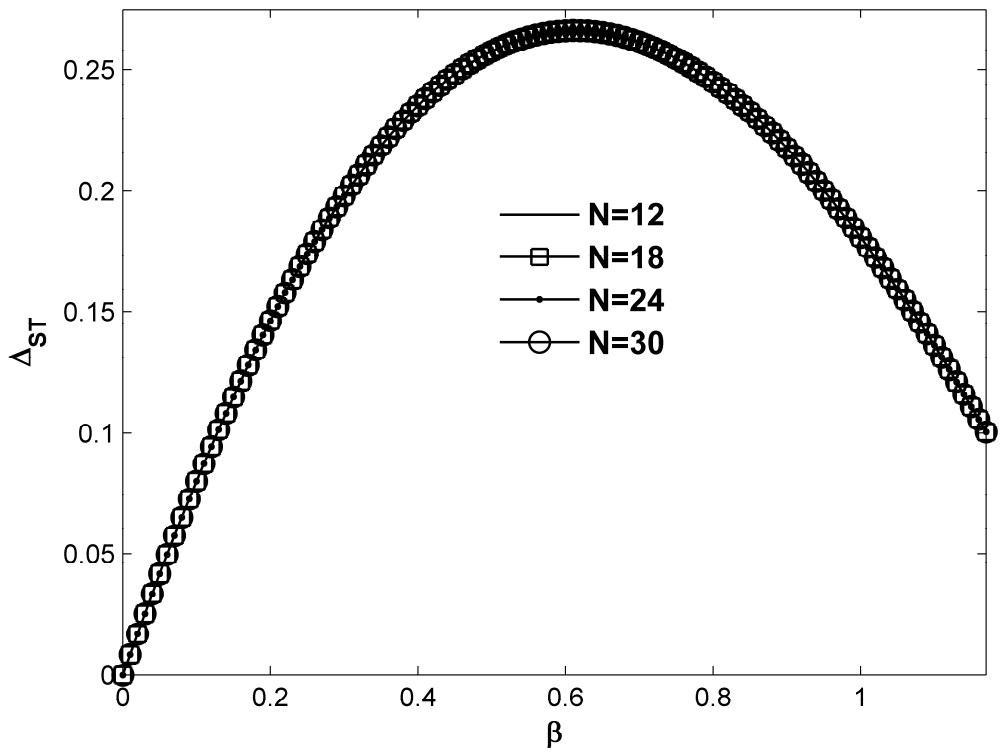

**Figure 9**

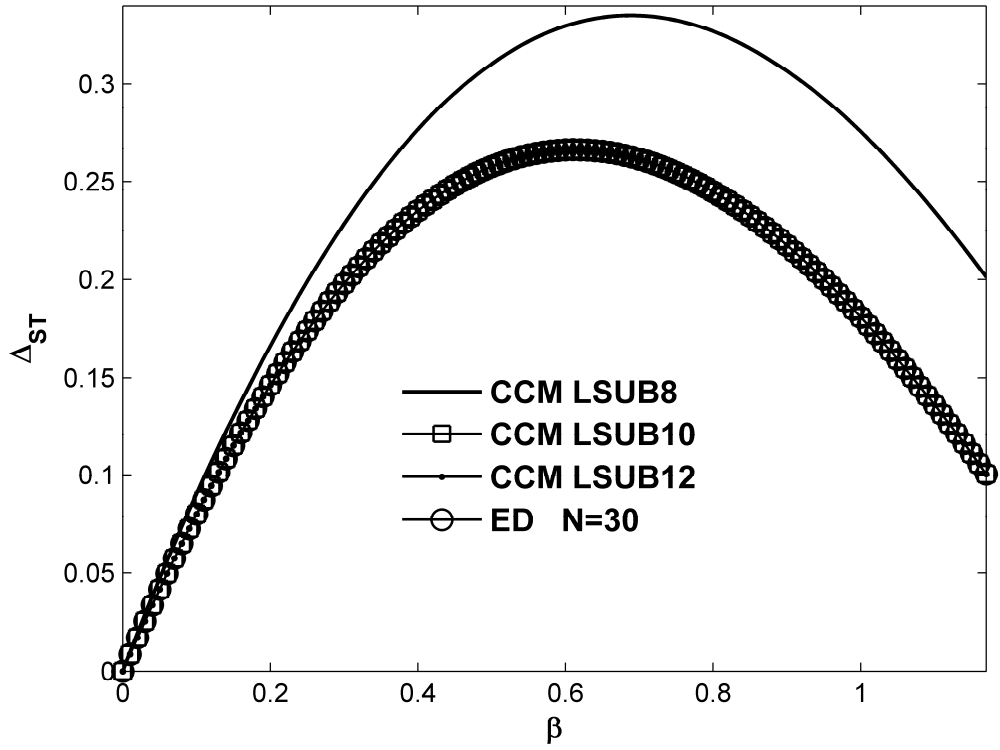

**Figure 10**

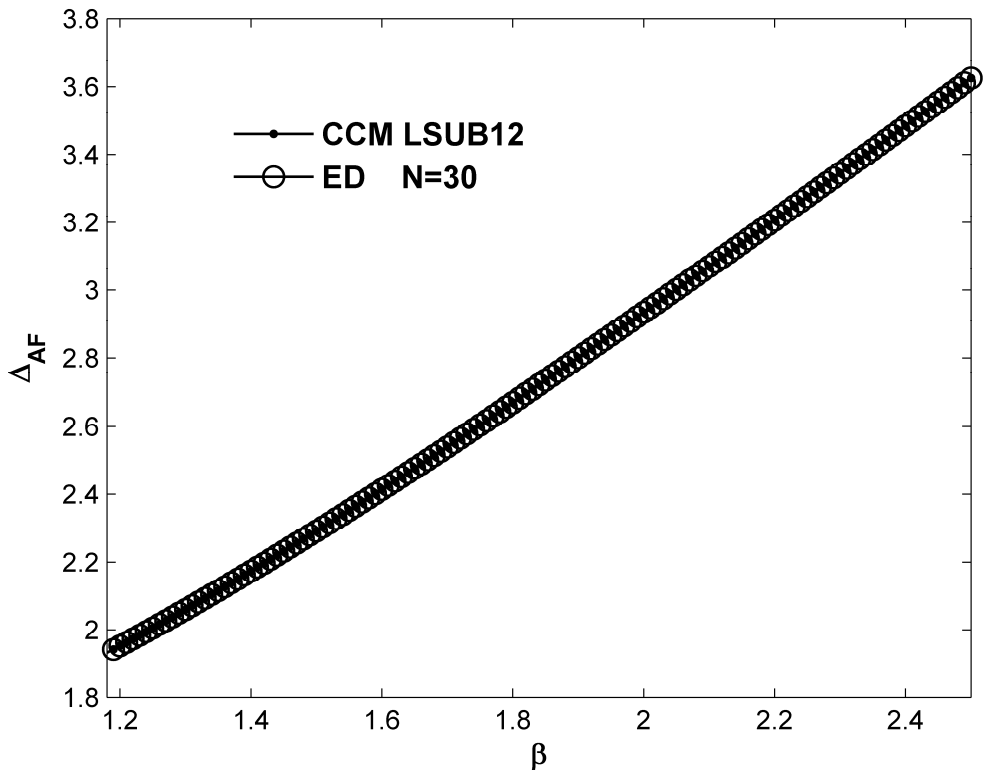